\documentclass{new_tlp}
\usepackage{mathptmx}

\newcommand\bcmdtab{\noindent\bgroup\tabcolsep=0pt%
  \begin{tabular}{@{}p{10pc}@{}p{20pc}@{}}}
\newcommand\ecmdtab{\end{tabular}\egroup}

\usepackage[latin1]{inputenc}

  \title[Theory and Practice of Logic Programming]
        {Indexing {\tt dif/2}}

  \author[U. Neumerkel and S. Kral]
         {ULRICH NEUMERKEL\\
         TU Wien, Vienna, Austria\\
         \email{ulrich@complang.tuwien.ac.at}
         \and STEFAN KRAL\\
         FH Wiener Neustadt, Austria\\
         \email{stefan.kral@fhwn.ac.at}}

\jdate{May 2016}
\pubyear{2016}
\pagerange{\pageref{firstpage}--\pageref{lastpage}}
\doi{n.a.}

\begin{document}
\maketitle

\begin{abstract}


Many Prolog programs are unnecessarily impure because of inadequate
means to express syntactic inequality.  While the frequently provided
built-in \verb`dif/2` is able to correctly describe expected answers, its
direct use in programs often leads to overly complex and inefficient
definitions --- mainly due to the lack of adequate indexing mechanisms.
We propose to overcome these problems by using a new predicate that
subsumes both equality and inequality via reification.  Code
complexity is reduced with a monotonic, higher-order if-then-else
construct based on \verb`call/N`. For comparable correct uses of impure
definitions, our approach is as determinate and similarly efficient as
its impure counterparts.

\end{abstract}

\begin{keywords}
Prolog, monotonicity, inequality, reification
\end{keywords}

\section{Introduction}

Do Prolog programmers really have to choose between logical purity and
efficiency?  Even for the most elementary notion of syntactic equality
this question still remains unanswered.  Today, many Prolog programs
consist of unnecessarily procedural constructs that have been
motivated by efficiency considerations blurring the declarative
vision.  To improve upon this situation we need pure constructs that
are of comparable efficiency as their impure counterparts.  We focus
our attention on the pure, monotonic subset of modern Prolog
processors.  The monotonic subset has many desirable properties: it
fits seamlessly with constraints, enables declarative debugging and
program slicing
techniques \cite{slice-use,gupu}, and is directly compatible with alternative search
procedures like iterative deepening.  Our effort aims into the same
direction that Functional Programming took so successfully --- away from
a command-oriented view to the pure core of the paradigm.

Many recent developments have facilitated pure programming in Prolog.
In particular, the widespread adoption of the higher-order built-in
predicate \verb`call/N` \cite{keefe1} together with its codification \cite{cor2} has
paved the way to yet unexplored pure programming
techniques.  On another track, more and more Prolog processors are rediscovering
the virtues of syntactic inequality via \verb`dif/2`.   

The very first Prolog, sometimes called
Prolog~0 \cite{prolog0} already supported \verb`dif/2`.
Unfortunately, the popular reimplementation Prolog~I \cite{prolog1-0} omitted
\verb`dif/2` and other coroutining features.  This
system was the basis for Edinburgh Prolog \cite{dec10-1978} which
led to ISO-Prolog \cite{std}.  After Prolog~I, \verb`dif/2` was
reintroduced in Prolog~II, independently reinvented in MU-Prolog
\cite{Naish}. and soon implementation schemes to integrate \verb`dif/2`
and coroutining into efficient systems appeared
\cite{Carlsson,Neumerkel}.  The major achievement was that the
efficiency of general Prolog programs {\em not} using \verb`dif/2`
remained unaffected within a system supporting \verb`dif/2`.  In this
manner \verb`dif/2` survived in major high-performance implementations
like SICStus.  However, it still has not gained general
acceptance among programmers.  We believe that the main reason for
this lack of acceptance is that
\verb`dif/2` does not directly deliver the abstraction that is
actually needed.  Its direct use leads to clumsy and unnecessarily
inefficient code.  Its combination with established control constructs
often leads to unsound results.  New, pure constructs are badly
needed.

\paragraph{Contents.} We first recall the deficiencies of
Prolog's if-then-else control constructs.  Then the
hidden deficiencies of the pure definition of \verb`member/2` are exposed.
A refined version is given whose efficiency is
subsequently improved with the help of reification and a new, pure and
monotonic if-then-else construct.  Finally, we show how our approach
permits to define more complex cases of reification
and compare it to constructive negation.

\section{The declarative limits of Prolog's if-then-else}

Prolog's if-then-else construct was first implemented in the
interpreter of DEC10 Prolog around 1978 \cite{dec10-1978}; its
compiler, however, did not support it.  Subsequent implementations,
starting with C-Prolog and Quintus Prolog, adopted it fully which led to its
inclusion into the ISO standard.

For many uses, this construct provides a clean way to express
conditional branching.  These uses all assume that the condition is
effectively stratified and {\em sufficiently instantiated} to permit
a simple test.  Some built-in predicates ensure their safe usage by
issuing instantiation errors in cases that are too general.  For
example, the built-in predicates for arithmetic evaluation and
comparison like \verb`(is)/2` and \verb`(>)/2` issue instantiation
errors according to the general scheme for errors \cite{error-k}. But in
general, problems arise.
For its common use, the construct \verb`( If_0 -> Then_0 ; Else_0 )`
contains three regular goals and is
equivalent to  \verb`( once(If_0) -> Then_0 ; Else_0 )`.  The first
answer of \verb`If_0` is taken, and all subsequent answers are
discarded.  The if-then-else has thus similar problems as a commit
operator.  Even the ``soft cut''-versions \verb`if/3` and
\verb`(*->)/2` of SICStus and SWI respectively, exhibit the same problems as Prolog's unsound negation.
MU-Prolog \cite{Naish} provided an implementation of if-then-else that
delays the goal \verb`If_0` until it is ground.  While sound, such an
implementation leads to many answers with unnecessarily floundering
goals.  Consider the goal \verb`[] = [E|Es]` which is not ground and
thus leads to floundering.  Even for the cases where this construct
works as expected, we still suffer from the lack of monotonicity.

\section{What's wrong with {\tt member/2}?}

Already pure definitions expose problematic behaviors that ultimately
lead to impure code.  Consider \verb`member/2`:

\begin{quote}
  \verb`member(X, L)` is true if \verb`X` is an element of the list \verb`L`.
\end{quote}

\noindent
The common actual definition is slightly more general than above
since \verb`L` is not required to be a list.  Certain
instances of partial lists like in the goal
\verb`member(a, [a|non_list])` succeed as well.  Such generalizations
are motivated by efficiency reasons; the cost for
visiting the entire list to ensure its well-formedness is often
not acceptable.  The complete definition of \verb`member/2` can thus be
described as:

\begin{quote}
  \verb`member(X, L)` is true iff \verb`X` is an element of a list prefix of \verb`L`.
\end{quote}

\begin{center}
\begin{verbatim}
member(X, [X|_Es]).
member(X, [_E|Es]) :-
   member(X, Es).

?- member(1, [1,2,3,4,5]).      ?- member(1, [1,2,1,4,5]).
   true                            true
;  false.                       ;  true
                                ;  false.
\end{verbatim}
\end{center}

\noindent
Above, lines starting with \verb`?-` show queries followed by their
answers---similar to SWI's top level shell.  Alternative answers are
separated by \verb`;`.  An alternative answer \verb`; false.`
indicates that Prolog needed further computation to ensure that no
further answer is present. It is thus an indication that Prolog still
uses space for this query, even though no further answer exists.  This
is thus a source of inefficiency we will address in this paper.

For its first answer the goal \verb`member(1, [1,2,3,4,5])` does not visit the
entire list.  Nevertheless, upon backtracking, the entire list gets
visited anyway.  Thus, the well-meant generalization does not lead to a
more efficient implementation.  For many goals with only a single
solution, \verb`member/2` leaves a choicepoint open that can only
be reclaimed upon failure or with non-declarative means like the cut.
So while \verb`member/2` is itself a pure definition,
its space consumption forces a programmer to resort to
impurity.  A common library predicate to this end is
\verb`memberchk/2` which does not leave any choicepoint open at the
expense of incompleteness.
The precise circumstances when this predicate is safe to use and
equivalent to \verb`member/2` are difficult to describe.  Many manuals suggest that the goal needs
to be {\em sufficiently instantiated} without giving a precise
criterion.  To err on the safe side, cautious programmers need to
add tests which are themselves prone to programming errors and incur
runtime overheads.  In the following example an
insufficiently instantiated goal leads to an unexpected failure.

\begin{center}
\begin{verbatim}
memberchk(X, Es) :-             ?- X = 2, memberchk(X, [1,2]), X = 2.
   once(member(X, Es)).            X = 2.    % expected solution

                                ?-        memberchk(X, [1,2]), X = 2.
                                   false.    % unexpected failure
\end{verbatim}
\end{center}

\section{Refurbishing {\tt member/2}}

The definition of \verb`member/2` contains unnecessary
redundancy.  This becomes more apparent when rewriting the two clauses to
an explicit disjunction.  In the first branch \verb`X = E` holds, but in the
second branch this may hold as well.  This can be observed with the
query \verb`member(1, [1,X])` shown below. The second answer
\verb`X = 1` is already subsumed
by the first answer \verb`true`.  The branches of the disjunction are thus not
mutually exclusive.  By adding an explicit \verb`dif/2`\footnote{An ISO conforming
  definition is given in the appendix for
  systems without {\tt dif/2} which resorts to instantiation errors in
  cases that cannot be handled correctly.} to the second
branch this redundancy is eliminated.  Note that there are still
possibilities for less-than-optimal answers as in the query
\verb`memberd(1, [X,1])` where the two answers could be merged into a
single answer.

\begin{center}
\begin{verbatim}
member(X, [E|Es]) :-            memberd(X, [E|Es]) :-
   (  X = E                        (  X = E
   ;  member(X, Es)                ;  dif(X, E),
   ).                                 memberd(X, Es)
                                   ).

?- member(1, [1,X]).            ?- memberd(1, [1,X]).
   true                            true
;  X = 1. % redundant answer    ;  false.

?- member(1, [X,1]).            ?- memberd(1, [X,1]).
   X = 1                           X = 1
;  true. % ~ redundant          ;  dif(X, 1)
                                ;  false.

                                ?- memberd(1, [1,2,3]).
                                   true
                                ;  false. % leftover choicepoint
\end{verbatim}
\end{center}

For sufficiently instantiated cases where \verb`memberchk/2` yields
correct results, there are no redundant answers for \verb`memberd/2`.
However, it still produces ``leftover choicepoints'' displayed as
\verb`; false`.  Space is thus consumed, even after succeeding.  This
is a frequent problem  when using \verb`dif/2` directly: it cannot help
to improve indexing since it is implemented as a separate built-in
predicate.  Indexing techniques have been developed both for the rapid
selection of matching clauses and to prevent the creation of
superfluous choicepoints.  They are even more essential to pure Prolog
programs which cannot resort to impure constructs like the cut.
With \verb`dif/2` in \verb`memberd/2` the situation is similar:
for the frequent case that \verb`X` and \verb`E` are
identical, a choicepoint is created even though we know that the goal
\verb`dif(X, E)` will fail upon backtracking.
Further, programming with \verb`dif/2` is rather
cumbersome since all conditions have to be stated twice: once for the
positive case and once for the negative.  Therefore, for both execution and
programmer efficiency, a new formulation is needed.

\section{Reification of equality}

The disjunction \verb`X = E ; dif(X, E)` is combined into a new
predicate \verb`=(X, E, T)` with an additional argument which is
\verb`true` if the terms are equal and \verb`false` otherwise.  In
this manner the truth value is reified.  An implementer is now free to
replace the definition of \verb`(=)/3` by a more efficient version.
The simple ISO conforming implementation in the appendix is already
able to eliminate many unnecessary choicepoints for all cases where
the terms are either identical or not unifiable.  A more elaborate
implementation might avoid to visit the terms several times.

\begin{center}
\begin{verbatim}
=(X, X, true).                  memberd(X, [E|Es]) :-
=(X, Y, false) :-                  =(X, E, T),
   dif(X, Y).                      (  T = true
                                   ;  T = false,
                                      memberd(X, Es)
                                   ).
\end{verbatim}
\end{center}

Still, this direct usage of reifying predicates does not address
all our concerns.  On the one hand there is an auxiliary variable for
each reified goal and on the other hand many Prolog implementations
cannot perform the above disjunction without a leftover choicepoint.  Both issues
are addressed using a higher-order predicate.

\section{The monotonic {\tt if\_/3}}

Our new, monotonic if-then-else is of the form \verb`if_(If_1, Then_0, Else_0)`.  The
condition \verb`If_1` is now no longer a goal but an incomplete
goal, which lacks one further argument. That argument is used for the
reified Boolean truth value.

\begin{center}
\begin{verbatim}
memberd(X, [E|Es]) :-           ?- memberd(1, [1,X]).
   if_( X = E       % (=)/3        true.   % fully deterministic
      , true
      , memberd(X, Es)          ?- memberd(1, [1,2,3]).
      ).                           true.   % fully deterministic
\end{verbatim}
\end{center}

\noindent
The implementation of \verb`if_/3` given in the appendix already
avoids many useless choicepoints.  Our choice to use a ternary
predicate in place of the nested binary operators was primarily
motivated by the semantic difficulties in ISO Prolog's if-then-else
construct whose principal functor is \verb`(;)/2` and
not \verb`(->)/2`.  This means that there are two entirely different
control constructs with the very same principal functor: 7.8.6 \verb`(;)/2` --
disjunction  and 7.8.8 \verb`(;)/2` -- if-then-else \cite{std}.  To avoid this
very hard-to-resolve ambiguity, we chose \verb`if_/3`.

\paragraph{Implementation.}
While our implementation of \verb`if_/3` in the appendix avoids the
creation of many useless choicepoints, the overall performance relies
heavily on the meta-call \verb`call/N`.  For this reason, we provide
an expanding version that removes many meta-calls in \verb`if_/3`.
This goal expansion is provided in \verb`library(reif)`.  Table
\ref{impure-vs-pure} compares two uses of \verb`memberchk/2` with
their pure counterparts.  First, the letter `z` is searched in a list
with all letters from a to z followed by a space.  The second test searches in a
list of pairs of the form \verb`Key-Value` the 10th element via a key (\verb`lassoc`).
Note that the impure and pure version are slightly different.
The impure version may skip invalid elements that are not pairs ;  the
pure version will fail in such a case.  The built-in \verb`memberchk/2`
and a version based on \verb`once(member(X,Xs))` are compared with
pure and complete versions, either using \verb`dif/2` directly ; or
using \verb`if_/3` directly ; as well as the expansion generated by \verb`library(reif)`.
The measurements were performed
with $10^6$ repetitions of the goals on an Intel Core i7-4700MQ 2.4 GHz.
We believe that the current overheads of a factor of 2 to 3 could be
further reduced if specialized built-ins for conditional
testing and a better register allocation scheme in SICStus were available.

\begin{table}[ht]

\caption{Runtimes of impure vs. pure definitions}
\label{impure-vs-pure}
\begin{minipage}{\textwidth}
\begin{tabular}{lcc|ccc}
\hline\hline
       &\multicolumn{2}{c|}{impure: {\tt memberchk/2}}
&\multicolumn{3}{c}{pure: {\tt memberd/2}} \\

system        & built-in  &  once   & dif     &    if\_/3   &  expanded \\ \hline
SICStus 4.3.2 &    0.480s &  0.480s &  1.070s &      4.610s & 0.900s    \\
SICStus 4.3.3 &    0.210s &  0.210s &  0.730s &      3.800s & 0.340s    \\
SWI 7.3.20    &    0.758s &  1.859s &  8.199s &      7.567s & 2.478s    \\
SWI 7.3.20 -O &    0.765s &  1.839s &  8.074s &      7.572s & 3.094s    \\
\hline
        &\multicolumn{2}{c|}{impure: \tt lassoc/3}     &\multicolumn{3}{c}{pure: \tt memberk/3} \\
system & built-in  & once                      & ---    & if\_/3   &
expanded     \\ \hline
SICStus 4.3.2 & 0.340s & 0.330s & &   2.110s &  0.480s \\
SICStus 4.3.3 & 0.190s & 0.200s & &   1.760s &  0.180s \\
SWI 7.3.20    & 0.585s & 1.102s & &   3.597s &  1.324s \\
SWI 7.3.20 -O & 0.827s & 1.555s & &   3.679s &  1.449s \\
\hline\hline
\end{tabular}
\end{minipage}
\end{table}

\newpage
Based on \verb`if_/3`, many idiomatic higher-order constructs can be
defined, now with substantially more general uses.  The commonly used
predicate for filtering elements of a list, often called
\verb`include/3` or \verb`filter/3`, is now replaced by a definition
\verb`tfilter/3` using a reified condition.  The first argument of \verb`tfilter/3` is
an incomplete goal which lacks two further arguments.  One for the
element to be considered and one for the truth value.  The following
queries illustrate general uses of this predicate that cannot be
obtained with the traditional definitions.

\begin{center}
\begin{verbatim}
tfilter(_CT_2,    [], []).
tfilter(CT_2, [E|Es], Fs0) :-
   if_(call(CT_2,E), Fs0 = [E|Fs], Fs0 = Fs ),
   tfilter(CT_2, Es, Fs).

?- tfilter(=(X), [1,2,3,2,3,3], Fs).
   X = 1, Fs = [1]
;  X = 2, Fs = [2,2]
;  X = 3, Fs = [3,3,3]
;  Fs = [], dif(X, 1), dif(X, 2), dif(X, 3).

duplicate(X, Xs) :-             ?- duplicate(X, [1,2,3,2,3,3]).
   tfilter(=(X), Xs, [_,_|_]).     X = 2
                                ;  X = 3
                                ;  false.
\end{verbatim}
\end{center}

\section{General reification}

So far, we have only considered the reified term equality predicate
\verb`(=)/3`.  Each new condition of \verb`if_/3` requires a new
reified definition.  In contrast to constructive negation
\cite{cnegchan,cnegdrabent}, these definitions are not constructed
automatically.  To test for membership the following reified
definition might be used.

\begin{center}
\begin{verbatim}
memberd_t(X, Es, true) :-
   memberd(X, Es).
memberd_t(X, Es, false) :-
   maplist(dif(X), Es).
\end{verbatim}
\end{center}

\noindent
This definition insists on a well-formed list in the negative case.
Contrast this to constructive negation which considers any failing
goal \verb`memberd(X, Es)` as a valid negative case.  Our definition thus
fails for \verb`memberd_t(X, non_list, T)`, that is, this case is
neither true nor false.  Constructive negation would consider this a
case for \verb`T = false`.  On the other extreme are approaches that
guarantee that type restrictions are maintained.  However, our
definition happens to be true for
\verb`memberd_t(1, [1|non_list], true)` --- for efficiency reasons.
A system maintaining the list-type would fail in this case whereas
we visit the list  only for the very necessary parts.
It is this freedom between the two extremes of constructive negation
and well-typed restrictions that permits an efficient implementation
of above definition.  In fact, the following improved definition
visits lists in the very same manner as the impure \verb`memberchk/2`
--- for comparable cases.  And for more general cases it
maintains correct answers.  The auxiliary \verb`l_memberd_t/3` permits
determinate list-traversal in systems with first-argument indexing only.

\begin{center}
\begin{verbatim}
memberd_t(X, Es, T) :-          l_memberd_t([], _, false).
   l_memberd_t(Es, X, T).       l_memberd_t([E|Es], X, T) :-
                                   if_( X = E
                                      , T = true
                                      , l_memberd_t(Es, X, T) ).

firstduplicate(X, [E|Es]) :-    ?- firstduplicate(1, [1,2,3,1]).
   if_( memberd_t(E, Es)           true.
      , X = E
      , firstduplicate(X, Es)   ?- firstduplicate(X, [1,2,2,1]).
      ).                           X = 1.

?- firstduplicate(X, [A,B,C]).
   X = A, A = B
;  X = A, A = C, dif(C, B)
;  X = B, B = C, dif(A, C), dif(A, C)
;  false.
\end{verbatim}
\end{center}

The following example shows how we deal with reification in the
general case.  Again, it guarantees a well-typed tree only for the
negative case.  In the positive case, the tree is visited only
partially.  Note that the reified version on the right uses a reified version of
disjunction.  Instead of \verb`(;)/2` the reified \verb`(;)/3` defined
in the appendix is used.  This reified version is now
significantly more compact than defining the positive and negative
cases explicitly.  In fact, it is close in size to the positive
case (\verb`treemember/2`) alone.

\newpage
\begin{center}
\begin{verbatim}
treemember_t(E, Tr, true) :-
   treemember(E, Tr).
treemember_t(E, Tr, false) :-
   tree_non_member(E, Tr).      treememberd_t(_, nil, false).
                                treememberd_t(E, t(F,L,R), T) :-
treemember(E, t(F,L,R)) :-         call(
   (  E = F                          (  E = F
   ;  treemember(E, L)               ;  treememberd_t(E, L)
   ;  treemember(E, R)               ;  treememberd_t(E, R)
   ).                                ),
                                   T).
tree_non_member(_, nil).
tree_non_member(E, t(F,L,R)) :-
   dif(E, F),
   tree_non_member(E, L),
   tree_non_member(E, R).
\end{verbatim}
\end{center}

\section{Conclusion}

We have presented a new approach to improving the efficiency of
pure programs using syntactic inequality.  Our solution to indexing
\verb`dif/2` was to provide a generalized reifying definition
together with a monotonic if-then-else construct.  In this manner the
calls to the actual built-in \verb`dif/2` are reduced to those cases
that actually need its general functionality.  In all other situations, the
programs run efficiently without calling \verb`dif/2` and without
creating many unnecessary choicepoints.

\paragraph{Acknowledgements.}
The presented programs were publicly developed on comp.lang.prolog
and as answers to questions on stackoverflow.com.

\begin{quote}

\noindent
   2009-10-15 ISO-\verb`dif/2` comp.lang.prolog
\\ 2012-12-01 Reification of term equality stackoverflow.com/q/13664870
\\ 2014-02-23 \verb`memberd/2` stackoverflow.com/a/21971885
\\ 2014-02-23 \verb`tfilter/3` stackoverflow.com/a/22053194
\\ 2014-12-09 \verb`if_/3` stackoverflow.com/a/27358600

\end{quote}

Further examples:
\verb`stackoverflow.com/search?q=[prolog]+if_`

\appendix

\section{Appendix}

The full \verb`library(reif)` and the benchmarks (\verb`memberbench`)
are available at: \\
\verb`http://www.complang.tuwien.ac.at/ulrich/Prolog-inedit/sicstus`

\begin{center}
\begin{verbatim}
dif(X, Y) :-
   X \== Y,
   ( X \= Y -> true ; throw(error(instantiation_error,_)) ).



% :- meta_predicate(if_(1, 0, 0)).
if_(If_1, Then_0, Else_0) :-
   call(If_1, T),
   (  T == true -> call(Then_0)
   ;  T == false -> call(Else_0)
   ;  nonvar(T) -> throw(error(type_error(boolean,T),_))
   ;  /* var(T) */ throw(error(instantiation_error,_))
   ).

=(X, Y, T) :-
   (  X == Y -> T = true
   ;  X \= Y -> T = false
   ;  T = true, X = Y
   ;  T = false,
      dif(X, Y)                             % ISO extension
      % throw(error(instantiation_error,_)) % ISO strict
   ).

','(A_1, B_1, T) :-
   if_(A_1, call(B_1, T), T = false).

;(A_1, B_1, T) :-
   if_(A_1, T = true, call(B_1, T)).
\end{verbatim}
\end{center}

\end{document}